\documentclass[aps,prl,twocolumn,showpacs,superscriptaddress]{revtex4}  
\usepackage{graphicx}  
\usepackage{dcolumn}   
\usepackage{bm}        
\usepackage{amssymb}   
\usepackage{rotating}

\newcommand{\TYNDALL}{Tyndall National Institute, Lee Maltings, University College Cork, Cork, Ireland} 
\newcommand{\MATHUCC}{School of Mathematical Sciences, University College Cork, Cork, Ireland}

\newcommand{\ddt}[1]{\frac{d #1}{dt}}

\begin{document}

\author{Nicholas Blackbeard}
\email{nicholas.blackbeard@tyndall.ie}
\affiliation{\TYNDALL}
\author{Simon Osborne}
\affiliation{\TYNDALL}
\author{Stephen O'Brien}
\affiliation{\TYNDALL}
\author{Andreas Amann}
\affiliation{\TYNDALL}
\affiliation{\MATHUCC}

\title{Bursting: when a cusp and a pitchfork interact}

\date{\today}

\begin{abstract}
We present an experimental and theoretical study of an unusual bursting mechanism in a two-mode semiconductor laser with single-mode optical injection. By tuning the strength and frequency of the injected light we find a transition from purely single-mode intensity oscillations to bursting in the intensity of the uninjected mode. We explain this phenomenon on the basis of a simple two-dimensional dynamical system, and show that the bursting in our experiment is organised by a cusp-pitchfork bifurcation of limit cycles.
\end{abstract}

\maketitle

Bursting is a striking feature in a wide range of physical and biological systems. It is characterised by periods of quiescence that are interspersed with periods of activity.  Well known systems that exhibit bursting include the Hodgkin--Huxley model \cite{Popovych2011,Li2010} and Taylor--Couette flows~\cite{Mullin1987,Abshagen2008}. 
Understanding the mechanisms responsible for bursting is a fundamental problem for dynamical systems theory and the physical sciences in general~\cite{Izhikevich2000,Golubitsky2001,Shilnikov2005}.

Systems with invariant manifolds are abundant in practical applications. Interesting dynamical phenomena are associated with the loss of transverse stability of an attractor contained in an invariant manifold~\cite{Ashwin1994,Ott2002}. For example, a blowout bifurcation leads to a form of bursting known as on-off intermittency. Here, the quiescent phase corresponds to chaotic dynamics confined to the invariant manifold. We recently verified experimentally that on-off intermittency occurs in an optically injected two-mode laser system~\cite{Osborne2012}.

In this letter we study theoretically and confirm experimentally a new mechanism that leads to bursting in the optically injected laser system. The mechanism is a global heteroclinic bifurcation which is organised by an interaction between a cusp and a pitchfork bifurcation of limit cycles. In contrast to on-off intermittency, the quiescent phase is governed by a nearby saddle-node bifurcation within the invariant manifold, and a transversally unstable limit cycle. The essential dynamics responsible for this bursting are captured by simple two-dimensional normal form equations for the cusp-pitchfork bifurcation of equilibria~\cite{Harlim2007}.

\begin{figure}
  \centering
  \includegraphics[width=\columnwidth]{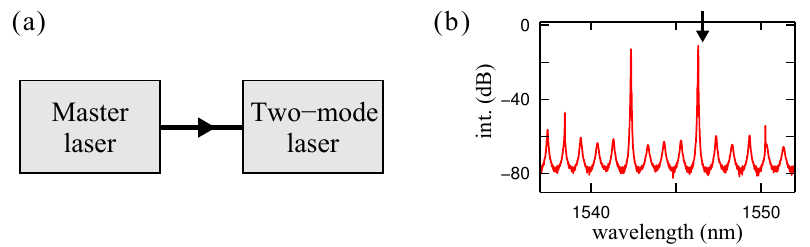}
\caption{\label{fig:setup} (Colour online) (a) Single-mode light from the tunable master laser is optically injected into the two-mode semiconductor laser. (b) Optical spectrum of the free running two-mode laser. The light is injected close to the long wavelength mode as indicated by the arrow. }
\end{figure}

\begin{figure}
\includegraphics[width=\columnwidth]{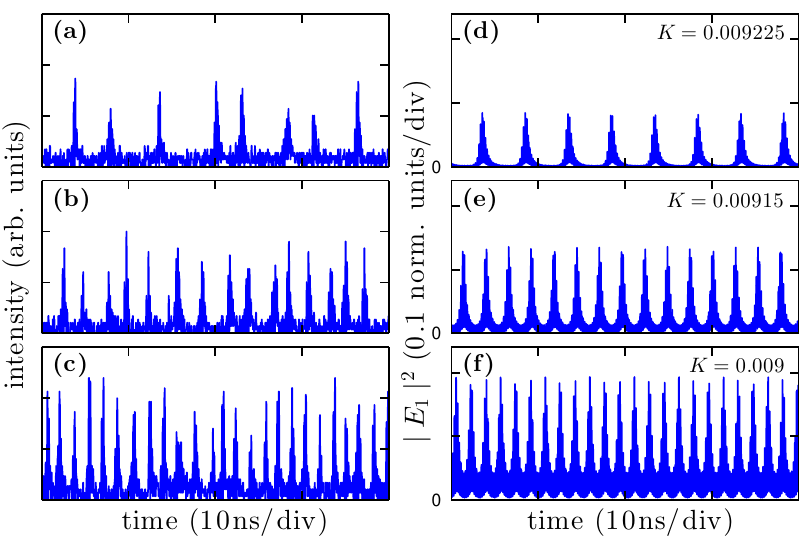}
\caption{\label{fig:bursting} Bursting time series of the uninjected mode obtained (a)--(c) experimentally and (d)--(f) theoretically. $\Delta\omega$ and $K$ take fixed values indicated by the white horizontal line in Fig.~\ref{fig:exp_and_num_bif_diagram}. From top to bottom the value of $K$ is decreased. }
\end{figure}

Our experiment employs a uni-directional coupling scheme as shown in Fig.~\ref{fig:setup}. 
The master laser is a narrow linewidth ($100 \ \mbox{kHz}$) tunable laser. The two-mode semiconductor laser is operated at twice threshold, where, in the absence of optical injection, weak coupling between the two modes leads to stable two-mode lasing (see the red optical spectra in Fig.~\ref{fig:setup}). Details of the design and lasing characteristics of this device are given in Ref.~\cite{Osborne2007}. Light emitted from the master laser is injected into the long wavelength mode of the two-mode laser. Depending on the strength and frequency of the injected light, drastic changes in the output of the two-mode laser can occur. These changes are visible in the time series of the individual modes which we recorded using high-speed photodiodes.

\begin{figure}
\includegraphics[width=\columnwidth]{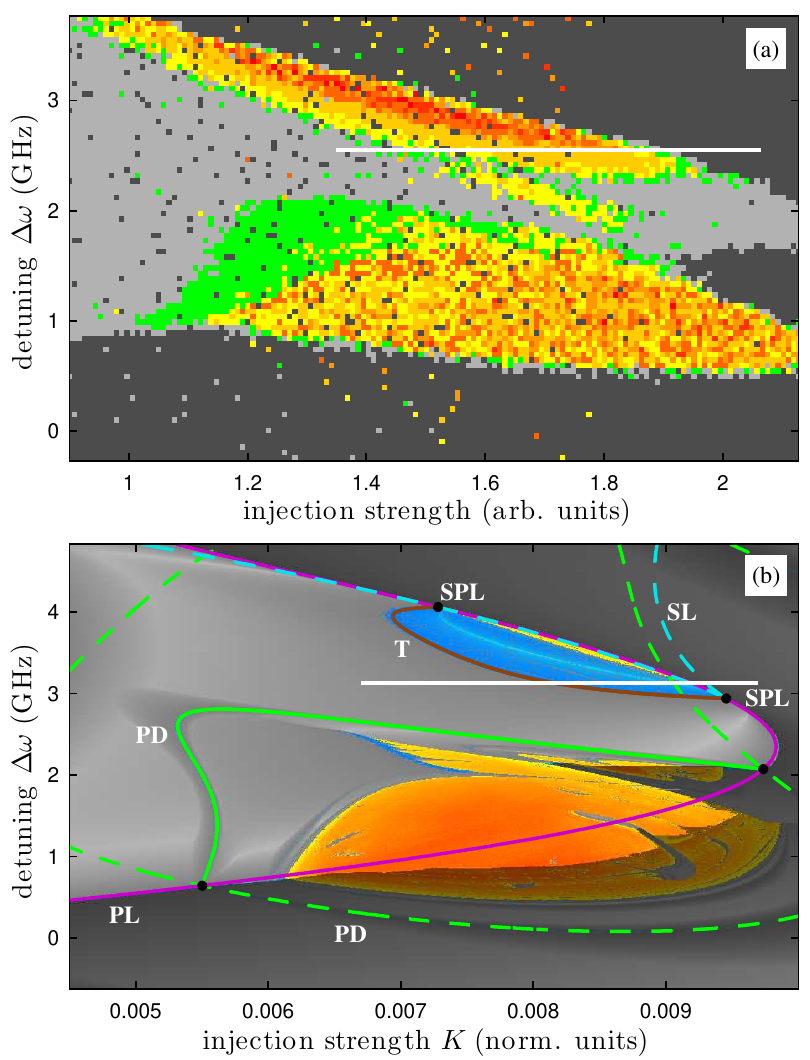}
\caption{\label{fig:exp_and_num_bif_diagram}(Colour online) Two parameter bifurcation diagrams. (a) experimental: colours represent the number of dominant peaks in the power spectrum of the uninjected mode, dark-grey$=0$, light-grey$=1$, green$=2$, and yellow--red$=3$--$9$. (b) numerics: bifurcations of limit cycles are saddle-node (SL), pitchfork (PL), period doubling (PD), torus (T), and saddle-node-pitchfork (SPL). Dashed lines indicate bifurcations solely of single-mode objects. Background colours indicate regions of periodic (grey), quasi-periodic (blue), and chaotic oscillations (yellow-red). Dark shading indicates that $|E_1(t)|=0$.}
\end{figure}

The bursting we are interested in occurs in a transition between single-mode dynamics, where the uninjected mode is suppressed and the intensity of the injected mode oscillates periodically, and two-mode dynamics, where both modes emit light with significant intensity. Figure~\ref{fig:bursting} shows time series of the uninjected mode during this transition for parameter values indicated by the white horizontal line in Fig.~\ref{fig:exp_and_num_bif_diagram}. A dramatic increase in the frequency of bursts is seen as the injection strength is decreased.

A simple four-dimensional set of ODEs reproduces this bursting behaviour (Fig.~\ref{fig:bursting}(d)--(f)), along with other experimentally observed phenomena~\cite{Osborne2009} for the setup shown in Fig.~\ref{fig:setup}. The normalised, time-dependent variables in this model are: the magnitude of the uninjected mode, $|E_1(t)|$, the complex field of the injected mode, $E_2(t)$, and the excess carrier density, $N(t)$. In dimensionless form it is given by~\cite{Osborne2009}: 
\begin{eqnarray}
  \ddt{|E_1|} &=& \frac{1}{2}\left((2N+1)g_1-1\right)|E_1| \, , \nonumber \\
  \ddt{E_2} &=&  \left[\frac{1}{2}((2N+1)g_2 -1)(1+i\alpha) -i\Delta\omega \right]E_2 + K \, , \nonumber \\
  T\ddt{N} &=& P - N - (1+2N)\left[ g_1 |E_1|^2 +  g_2 |E_2|^2 \right]\, , \label{system}
\end{eqnarray} 
where time is measured in units of the photon lifetime, $\tau_{ph}=1/(9.8\times10^{11}) \ \mbox{s}$. The bifurcation parameters are the normalised injection strength, $K$, and the normalised frequency difference (or detuning) between the injected mode and the injected light, $\Delta\omega$. Further parameters are the coupling, $\alpha$, between the magnitude and phase of $E_2$; the product of carrier lifetime and the cavity decay rate, $T$; and the normalised pump current, $P$. The two modes are coupled through $N(t)$ via the nonlinear modal gain
\begin{eqnarray*}
  g_1=\left[1+ \epsilon \left(|E_1|^2 + \beta |E_2|^2 \right)\right]^{-1} \, ,\\
  g_2=\left[1+ \epsilon \left(\beta |E_1|^2 + |E_2|^2 \right)\right]^{-1} \, ,
\end{eqnarray*}
where $\epsilon$ is the self saturation of a mode, and  $\epsilon \beta$  is the cross saturation between modes. To agree with the experiments in Ref.~\cite{Osborne2009} we fix $\epsilon=0.01$,  $\beta=2/3$, $\alpha=2.6$, $T=800$, and $P=0.5$ (twice threshold).

The phase space of system~(\ref{system}) features a three-dimensional sub-manifold, 
$$\mathcal{M}=\{ \ \ (|E_1|, \mbox{Re}(E_2), \mbox{Im}(E_2),N)\ \in \mathbb{R}^4 \  : \ \ |E_1|=0 \ \ \} \, ,$$ 
which is invariant under the time evolution of (\ref{system}). We refer to $\mathcal{M}$ as the {\it single-mode manifold} since the dynamics on $\mathcal{M}$ reduce to that of the well-studied single-mode injection problem~\cite{Wieczorek2005b,Kelleher2011}. System~(\ref{system}) has a formal $\mathbb{Z}_2$ symmetry due to the transformation $|E_1| \to -|E_1|$. Only the positive magnitudes are relevant but the $\mathbb{Z}_2$ symmetry alters the generic set of bifurcations for system~(\ref{system}).

Figure~\ref{fig:exp_and_num_bif_diagram} contains two-parameter bifurcations diagrams in the $(K,\Delta \omega)$-plane. In the experimental bifurcation diagram (Fig.~\ref{fig:exp_and_num_bif_diagram}(a)) regions of the parameter plane are coloured to represent the number of dominant peaks in the power spectrum of the uninjected mode for frequencies less than 1.5 times the relaxation oscillation. Figure~\ref{fig:exp_and_num_bif_diagram}(b) contains a partial bifurcation diagram for system~(\ref{system}). Curves are bifurcations continued using AUTO~\cite{Doedel2007} and the coloured background represents different attractor types characterised by their Lyapunov spectrum~\cite{Benettin1980}. Additionally, parameter regions are shaded if the corresponding attractor is contained in the single-mode manifold $\mathcal{M}$ (i.e. $|E_1(t)|=0$). The qualitative agreement between experiment and theory is clear. The dark-grey region in Fig.~\ref{fig:exp_and_num_bif_diagram}(a) corresponds to the uninjected mode being off. Leaving the dark-grey region, e.g. by increasing $\Delta\omega$, results in the uninjected mode turning on. Exactly how the uninjected mode turns on depends on where the dark-grey region's boundary is crossed. On the boundary between dark- and light-grey are supercritical pitchfork of limit cycle bifurcations (Fig.~\ref{fig:exp_and_num_bif_diagram}(b)). Crossing this boundary causes the uninjected mode to gradually turn on. There are two boundaries where the uninjected mode turns on in a non-trivial manner. Both of these boundaries are associated with bursting dynamics. In the experimental bifurcation diagram (Fig.~\ref{fig:exp_and_num_bif_diagram}(a)), these are the boundaries between dark-grey and yellow/red regions. On-off intermittency was  found along the boundary at low $\Delta\omega$ \cite{Osborne2012}. In this letter we focus on the second boundary at larger $\Delta\omega$. On this boundary there are two saddle-node pitchfork of limit cycle (SPL) bifurcations. These are formed by tangencies between a curve of saddle-node of limit cycle bifurcations and a curve of pitchfork of limit cycle bifurcations (Fig.~\ref{fig:exp_and_num_bif_diagram}(b)). Connecting the two SPL points is a curve of supercritical torus bifurcations. As we will show, this arrangement of bifurcations, which we refer to as the torus bubble, is due to a nearby interaction between a cusp and a pitchfork bifurcation of limit cycles.

\begin{figure}
\includegraphics[width=0.8\columnwidth]{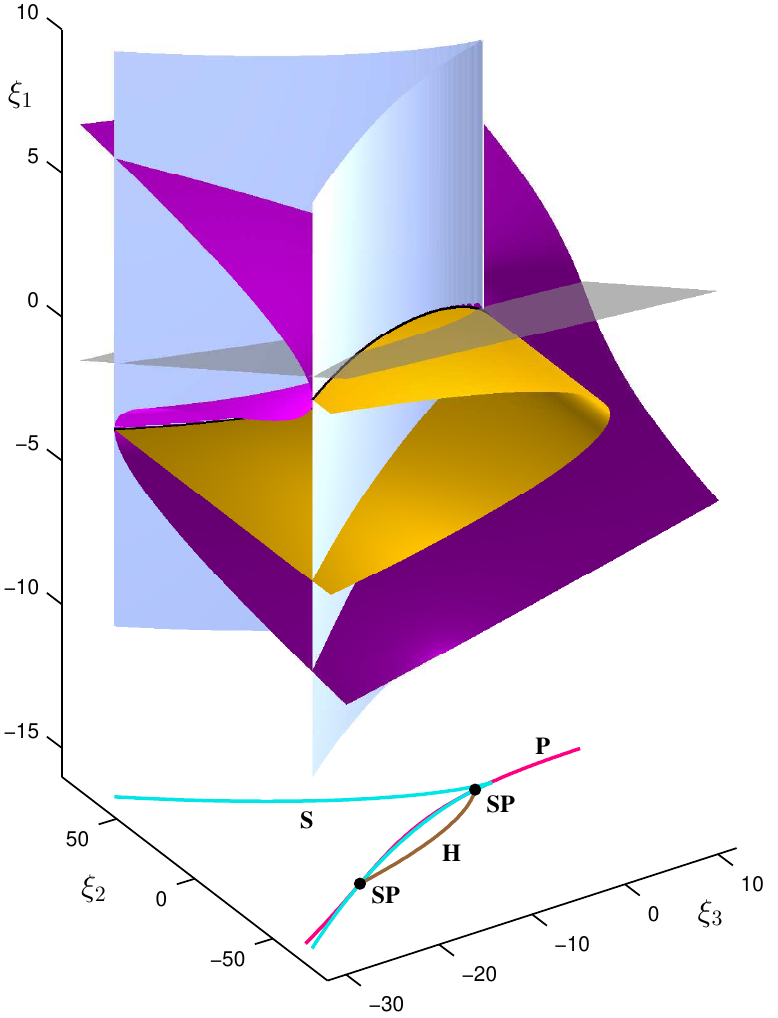}
\caption{\label{fig:unfolding_surfaces}(Colour online) Unfolding of a cusp-pitchfork bifurcation. Saddle-node surface (blue), pitchfork surface (purple), Hopf surface (brown), and saddle-node-pitchfork curve (black). The bifurcation curves on the bottom are projections of the intersection of the transparent surface with the bifurcations. Bifurcations of equilibria are saddle-node (S), pitchfork (P), Hopf (H), and saddle-node pitchfork (SP). }
\end{figure}

In order to gain analytic insight into the bifurcation structure, we first study a simplified system where the limit cycles are replaced by equilibria.  We demonstrate that a large part of the phenomena observed in Fig.~3, including the bursting shown in Fig. 2, are a consequence of the unfolding of a cusp-pitchfork bifurcation~\cite{Harlim2007}.  Similar to Ref.~\cite{Harlim2007} we choose a simple two-dimensional normal form for a system which is close to a cusp-pitchfork bifurcation,
\begin{eqnarray}
  \dot{r} &=& r(\xi_1 + z) \, ,\nonumber \\
  \dot{z} &=& -\xi_2 - \xi_3 z - r^2 - z^3 \, . \label{eq:cusp_pitchfork_normal_form} 
\end{eqnarray}
The bifurcation parameters are $\xi_1,\xi_2,\xi_3 \in \mathbb{R}$ and we note that the phase space $(r,z)$ of system~(\ref{eq:cusp_pitchfork_normal_form}) has an invariant manifold defined by $r=0$, analogous to the single-mode manifold $\mathcal{M}$ of system~(\ref{system}). The local bifurcation structure of system~(\ref{eq:cusp_pitchfork_normal_form}) is shown in Fig.~\ref{fig:unfolding_surfaces}. Saddle-node bifurcations take place on the (blue) surface that resembles an open book. For parameter values on the outside of the open book ($\xi_3 > -3(\xi_2/2)^{2/3}$), system~(\ref{eq:cusp_pitchfork_normal_form}) has one equilibrium in the invariant manifold. For parameters values inside the open book ($\xi_3 < -3(\xi_2/2)^{2/3}$), system~(\ref{eq:cusp_pitchfork_normal_form}) has three equilibria in the invariant manifold. The (purple) folded surface in Fig.~\ref{fig:unfolding_surfaces} corresponds to pitchfork bifurcations. Below the pitchfork surface system~(\ref{eq:cusp_pitchfork_normal_form}) has no equilibria off the invariant manifold. Crossing through the pitchfork surface causes two symmetrically related equilibria to bifurcate out of the invariant manifold; one with $r>0$ and the other with $r<0$. The remaining (brown) surface corresponds to Hopf bifurcations where each of the symmetrically related equilibria bifurcate to a limit cycle. The three surfaces meet at the black curve which comprises two branches of saddle-node-pitchfork bifurcations that emanate from the cusp-pitchfork bifurcation at the origin.

\begin{figure}
\includegraphics[width=0.8\columnwidth]{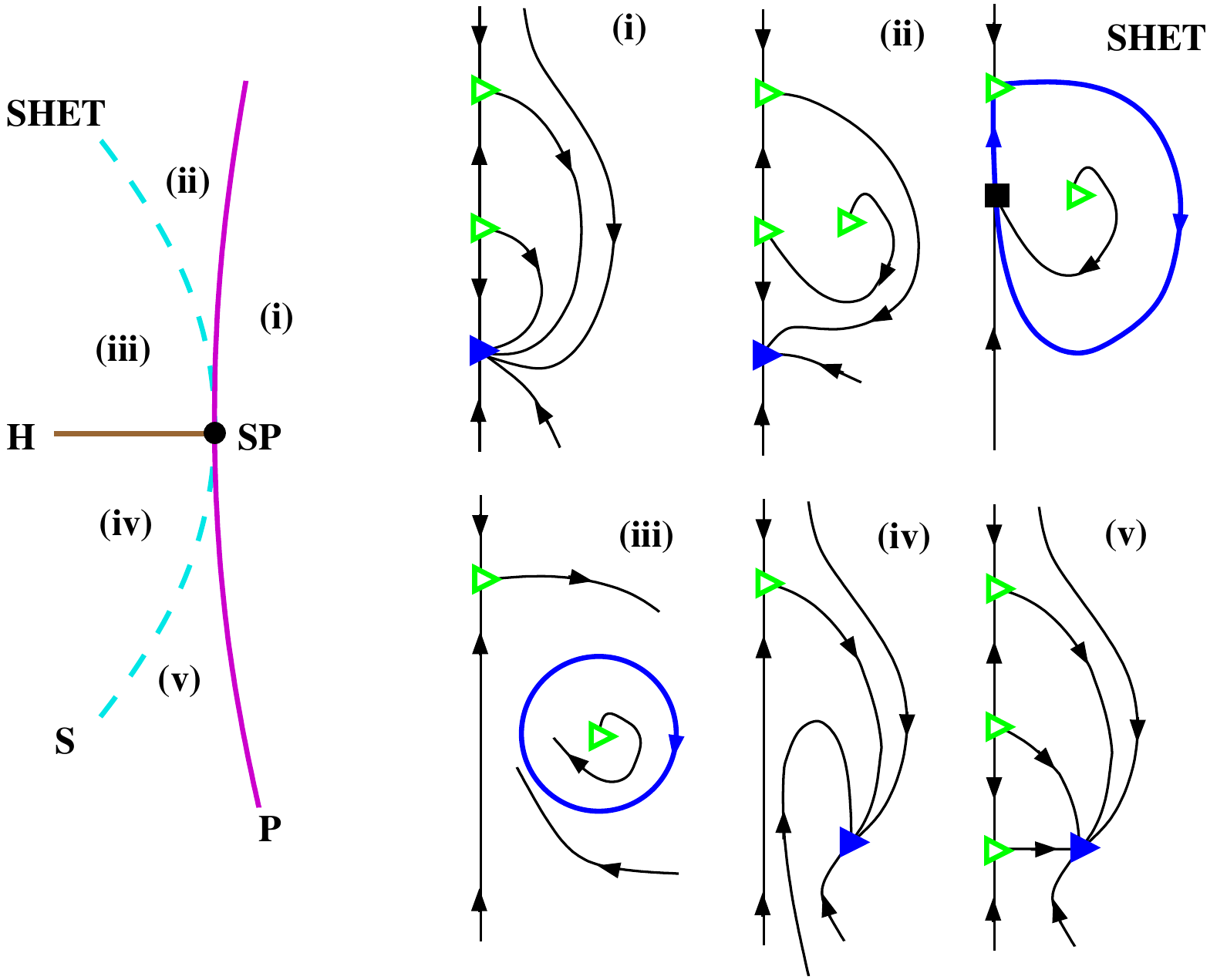}
\caption{\label{fig:normal_form_phase_portraits}(Colour online) Bifurcation diagram for the saddle-node-pitchfork bifurcation. (i)--(v) phase portraits in the $(r,z)$ plane. The vertical lines corresponds to the invariant manifold ($r=0$). Closed triangle$=$stable equilibrium, open triangle$=$saddle-equilibrium, and the closed square$=$nonhyperbolic equilibrium.}
\end{figure}

The transparent surface in Fig.~\ref{fig:unfolding_surfaces} indicates a slice of parameter space. Intersections of the bifurcations with this surface are projected down to the bottom of the figure. From these projections we see the same arrangement of bifurcations as in the torus bubble (Fig.~\ref{fig:exp_and_num_bif_diagram}), but for equilibria instead of limit cycles.

Figure~\ref{fig:normal_form_phase_portraits} contains a sketch of the phase portraits around the saddle-node pitchfork (SP) point closest to the cusp-pitchfork bifurcation. We focus on the curve of global saddle-node heteroclinic (SHET) bifurcations. Along the SHET bifurcation curve, system~(\ref{eq:cusp_pitchfork_normal_form}) has a transversally stable, nonhyperbolic equilibrium; and a heteroclinic cycle connecting the nonhyperbolic equilibrium to a transversally unstable saddle equilibrium. For parameter values on one side of the SHET curve (region (ii)) the nonhyperbolic equilibrium splits into two equilibria and the heteroclinic cycle is destroyed. For parameter values on the other side of the SHET curve (region (iii)) the nonhyperbolic equilibrium is destroyed and the heteroclinic cycle bifurcates to a stable limit cycle. In region (iii), close to the SHET curve, the ghost of the saddle-node bifurcation results in a slow region of phase space. While a trajectory of system~(\ref{eq:cusp_pitchfork_normal_form}) is in this slow region it is attracted towards the invariant manifold (quiescent phase). Once through the slow region the trajectory shadows the dynamics in the invariant manifold, and is thus attracted toward a saddle equilibrium which is stable within the invariant manifold but transversally unstable. This equilibrium repels the trajectory away from the invariant manifold (active phase) and the trajectory is then globally reinjected via the limit cycle to the slow region of phase space for the process to start over again.

\begin{figure}
\includegraphics[width=\columnwidth]{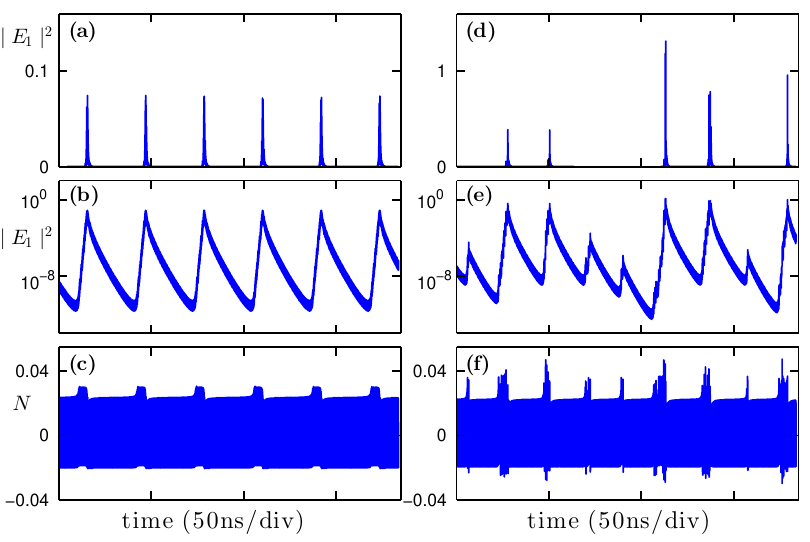}
\caption{\label{fig:periodic_vs_chaotic}(a)--(c) periodic bursting for $K=0.0092473$ and $\Delta\omega=3.088 \ \mbox{GHz}$. (d)--(f) chaotic bursting for $K=0.0085$ and $\Delta\omega=3.5126 \ \mbox{GHz}$.}
\end{figure}

The mechanism responsible for bursting in the two-mode laser (Fig.~\ref{fig:bursting}) is analogous to that described in the previous paragraph but for limit cycles instead of equilibria. We illustrate this in Fig.~\ref{fig:periodic_vs_chaotic}(a)--(c). Figure~\ref{fig:periodic_vs_chaotic}(a) shows bursting in the time series of the uninjected mode. The same plot on a log scale (Fig.~\ref{fig:periodic_vs_chaotic}(b)) reveals an almost linear trend towards the single-mode manifold $\mathcal{M}$. The slope of this linear trend is related to the negative transverse Lyapunov exponent of the nearby saddle-node limit cycle. Similarly, there is a phase of linear trend away from $\mathcal{M}$. The slope of this trend is related to the positive transverse Lyapunov exponent of the saddle-cycle involved in the nearby SHET bifurcation. Moreover, two distinct periodic-like oscillations can be seen in the time series of the laser's population inversion (Fig.~\ref{fig:periodic_vs_chaotic}(c)). The time-scale of the quiescent phase is strongly influenced by the slow region of phase space. Since the slow region is caused by the ghost of a saddle-node of limit cycle we can estimate that the quiescent phase scales as $\mu^{-\frac{1}{2}}$, where $\mu$ is the distance from the saddle-node of limit cycle bifurcation. We were able to experimentally verify this scaling.

While the bifurcations responsible for the bursting in Fig.~\ref{fig:periodic_vs_chaotic}(a)--(c) are similar to those in system~(\ref{eq:cusp_pitchfork_normal_form}), the cusp-pitchfork bifurcation in the two-mode laser (Fig.~\ref{fig:exp_and_num_bif_diagram}) is of limit cycles, not equilibria. The extra rotation can complicate the bifurcation structure. For example, the stable torus could break up before the saddle-node of limit cycle bifurcation, and any curves of global bifurcations could be replaced by cones of heteroclinic tangles~\cite{Kirk1991,Champneys2004}. Furthermore, additional bifurcations can occur away from the cusp-pitchfork bifurcation. These complications can lead to different types of bursting. Such an example is shown in Fig.~\ref{fig:periodic_vs_chaotic}(d)--(f). Here, the transversally unstable saddle-cycle is replaced by a chaotic saddle whose unstable direction is transverse to $\mathcal{M}$. The bursts are not (quasi-)periodic and the amplitude of the bursts varies from burst to burst (Fig.~\ref{fig:periodic_vs_chaotic}(d)). The same plot on a log scale (Fig.~\ref{fig:periodic_vs_chaotic}(e)) reveals phases of a linear trend towards $\mathcal{M}$, again this is associated with the negative transverse Lyapunov exponent of the nearby saddle-node limit cycle. The diffusive nature of the growth away from $\mathcal{M}$ is strongly indicative of a transversally unstable (to $\mathcal{M}$) chaotic saddle. When a trajectory is very close to ${M}$, a longer signature of the chaotic saddle is observed  before bursting away from $\mathcal{M}$ (Fig.~\ref{fig:periodic_vs_chaotic}(f)).

In summary, we identified the cusp-pitchfork bifurcation of limit cycles as a mechanism for experimentally observed bursting in a two-mode semiconductor laser with optical injection. The backbone of the bifurcation structure responsible for this bursting was explained by simple two-dimensional normal form equations for the cusp-pitchfork bifurcation of equilibria.

This work was supported by Science Foundation Ireland. We thank Eblana Photonics for the fabrication of the two-mode laser.

\end{document}